\documentclass{PoS}

\newcommand{\Tr}{\mathsf{Tr}}

\title{'t Hooft loop and the phases of SU(2) LGT}

\ShortTitle{'t Hooft loop and the phases of SU(2) LGT}

\author{\speaker{G.~Burgio}\\
        Institut f\"ur Theoretische Physik\\
        Auf der Morgenstelle 14\\
        72076 T\"ubingen\\
        Germany\\
        E-mail: \email{giuseppe.burgio@uni-tuebingen.de}}

\abstract{We analyze the vacuum structure of SU(2) lattice gauge theories in 
$D=2,3,4$, concentrating on the stability of 't Hooft loops. High precision 
calculations have been performed in D=3; similar results 
hold also for D=4 and D=2. We discuss the impact of our findings on the 
continuum limit of Yang-Mills theories.}

\FullConference{31st International Symposium on Lattice Field Theory - 
                LATTICE 2013\\
		July 29 - August 3, 2013\\
		Mainz, Germany}

\begin{document}

\section{Introduction}

Most of the ideas aimed at solving confinement in $SU(N)$ Yang-Mills theories 
involve topological degrees of freedom of some sort. Among these, 
$\mathbb{Z}_N$ (i.e. center) vortices \cite{'tHooft:1979uj} have received much 
attention in the literature in general and in lattice investigations in 
particular \cite{Del Debbio:1996mh,Langfeld:1997jx}.

The gauge group of pure Yang-Mills $SU(N)/\mathbb{Z}_N$ possesses a non trivial 
first homotopy class, corresponding to its center, 
$\pi_1(SU(N)/\mathbb{Z}_N) = \mathbb{Z}_N$. A super-selection rule will thus
arise for the physical Hilbert space of gauge invariant states 
\cite{Burgio:1999tg}, with sectors 
labeled by a center vortex topological index $n \in \mathbb{Z}_N$. 
According to 't Hooft's original idea \cite{'tHooft:1979uj}, the low 
temperature confinement phase should then correspond to a superposition of all 
topological sectors, while above the deconfinement transition vortex symmetry 
gets broken to the trivial sector $n = 0$; the 't Hooft loop $H$, dual
to the Wilson loop $W$, is the 
natural observable to describe the transition, ``counting'' the number of 
topological vortices piercing it. From $H$ one can reconstruct the free energy 
for vortex 
creation, $F = \Delta U -T \Delta S$, which should jump at $T_c$; the 
monitoring of such behaviour across the deconfinement transition has received 
broad attention in the literature
\cite{Kovacs:2000sy,deForcrand:2000fi,deForcrand:2001nd,Burgio:2006dc,%
Burgio:2006xj}.

The natural choice to investigate $F$
upon lattice discretization of Yang-Mills theories would be to define the
partition function through the adjoint 
Wilson action $Z \sim e^{\beta_A \Tr_A(U)}$, transforming under 
$SU(N)/\mathbb{Z}_N$; 
in this case all 
topological sectors are dynamically included \cite{deForcrand:2002vs}. 
Universality should of course allow the equivalent use of the standard Wilson 
plaquette action $S\sim\Tr_F(U)$. In this case topology must however be 
introduced ``by hand'' summing over all twisted boundary
conditions\footnote{Such topological boundary conditions also play a r\^ole in 
lattice investigations of the string spectrum or large $N$ reduction.}. 
The ``proper'' partition function $\tilde{Z}$ can then be 
defined through the weighted sum of all partition functions with fixed 
twisted b.c.\footnote{See e.g. Ref.~\cite{vonSmekal:2012vx}, Chapt.~3.}. Since 
each of them must be determined by 
independent simulations, their relative weights can only be calculated through 
indirect means \cite{deForcrand:2000fi,deForcrand:2001nd,vonSmekal:2012vx}. 

There is however a loophole in the argument given above. The two partition 
functions $Z$ and $\tilde{Z}$ can only 
be shown to be equivalent when $\mathbb{Z}_N$
magnetic monopoles are absent \cite{deForcrand:2002vs,Mack:1978rq}. Taking the 
explicit case of $SU(2)/\mathbb{Z}_2 = SO(3)$, this translates into the 
constraint:
\begin{equation}
\sigma_c=\prod_{\overline{P}\in\partial c}\mathrm{sign}
(\Tr_{F} \,U_{\overline{P}})=1
\label{eq:mon}
\end{equation}
being satisfied for every elementary 
3-cube $c$, where $U_{\overline{P}}$ denotes the plaquettes belonging to the cube 
surface $\partial c$. This ensures that endpoints of open center vortices, 
${\mathbb{Z}}_2$ magnetic monopoles, are 
suppressed and only {\it closed}, i.e. topological ${\mathbb{Z}}_2$ vortices 
winding around the boundaries can form. 

The above condition is usually quoted when claiming that the bulk transition 
separating the strong and weak coupling regime along $\beta_A$ 
\cite{Greensite:1981hw,Bhanot:1981eb,Halliday:1981te,Halliday:1981tm,%
Baig:1987qa} constitutes an obstacle in defining the continuum limit for the 
adjoint Wilson action. This however assumes that topological sectors along 
the fundamental coupling
$\beta$ are always well defined. We will show this {\it not} to be the case. 
Together with a set of established results demonstrating that
above the adjoint bulk transition topological sectors are well defined and
a physical continuum limit of the theory exists \cite{Burgio:2006dc,%
Burgio:2006xj,Barresi:2001dt,Barresi:2002un,Barresi:2003jq,Barresi:2003yb,%
Barresi:2004qa,Barresi:2004gk,Burgio:2005xe,Barresi:2006gq}, we can turn the 
argument around, casting doubts that investigations of vortex topology for the 
fundamental Wilson are well defined. Preliminary results had been presented in 
Ref.~\cite{Burgio:2007np}.

\section{Setup}

We will investigate as a test case the $SU(2)$ fundamental Wilson action with 
periodic b.c.:
\begin{equation}
S = \beta \sum_{x,\mu>\nu}[1-\Tr_F(U_{\mu\nu}(x)]
\end{equation}
in $D=2,3,4$ dimensions. Different groups or b.c. can be considered as well and
won't change the main results given below.

The twist operator, measuring the number of topological vortices piercing 
the 't Hooft loop in the $\mu,\nu$ planes, can be constructed via 
\cite{Kovacs:2000sy}:
\begin{equation}
z_{\mu\nu} = \frac{1}{L^{D-2}}\,\sum_{\vec{y}\bot{\mu\nu}{\rm plane}}\; \prod_{\vec{x} \in {\mu\nu} {\rm plane}} 
{\sf sign}({\Tr}_{F}U_{\mu\nu}(\vec{x},\vec{y}))\,.
\label{eq:twist}
\end{equation}
When topological sectors are well defined $z_{\mu\nu}$ takes 
values $\pm 1$ for all fixed $\mu$ and $\nu$; e.g. in the case at hand, i.e.
periodic b.c., the topological sector must be trivial and one should always have 
$z_{\mu\nu}=1$ $\forall \mu, \nu$.
It is now easy to define an order parameter $z$ such that $z=1$ if, whatever 
the b.c., the vortex topology takes the correct value expected in the 
continuum theory, while $z=0$ when $\mathbb{Z}_2$ 
monopoles are still present and open $\mathbb{Z}_2$ center vortices dominate 
the vacuum, making the identification of topological sectors ill defined at 
best:
\begin{eqnarray}
z &=& 1-| z_{12} - \langle z_{12} \rangle|\qquad\qquad\qquad  D=2\\
z &=&\frac{2}{D(D-1)} \sum_{\mu>\nu=1}^{D} \langle |z_{\mu\nu}|\rangle \quad\qquad D \geq 3\,.
\end{eqnarray}
In the following we will monitor $z$ as a function of $\beta$ and the volume 
$L^D$. The continuum limit of our lattice discretization is of course
defined by taking the thermodynamic limit $L\to\infty$ first and then
the weak coupling limit $\beta\to\infty$.

\section{Results}

\begin{figure}
\begin{center}
\includegraphics[width=0.49\textwidth]{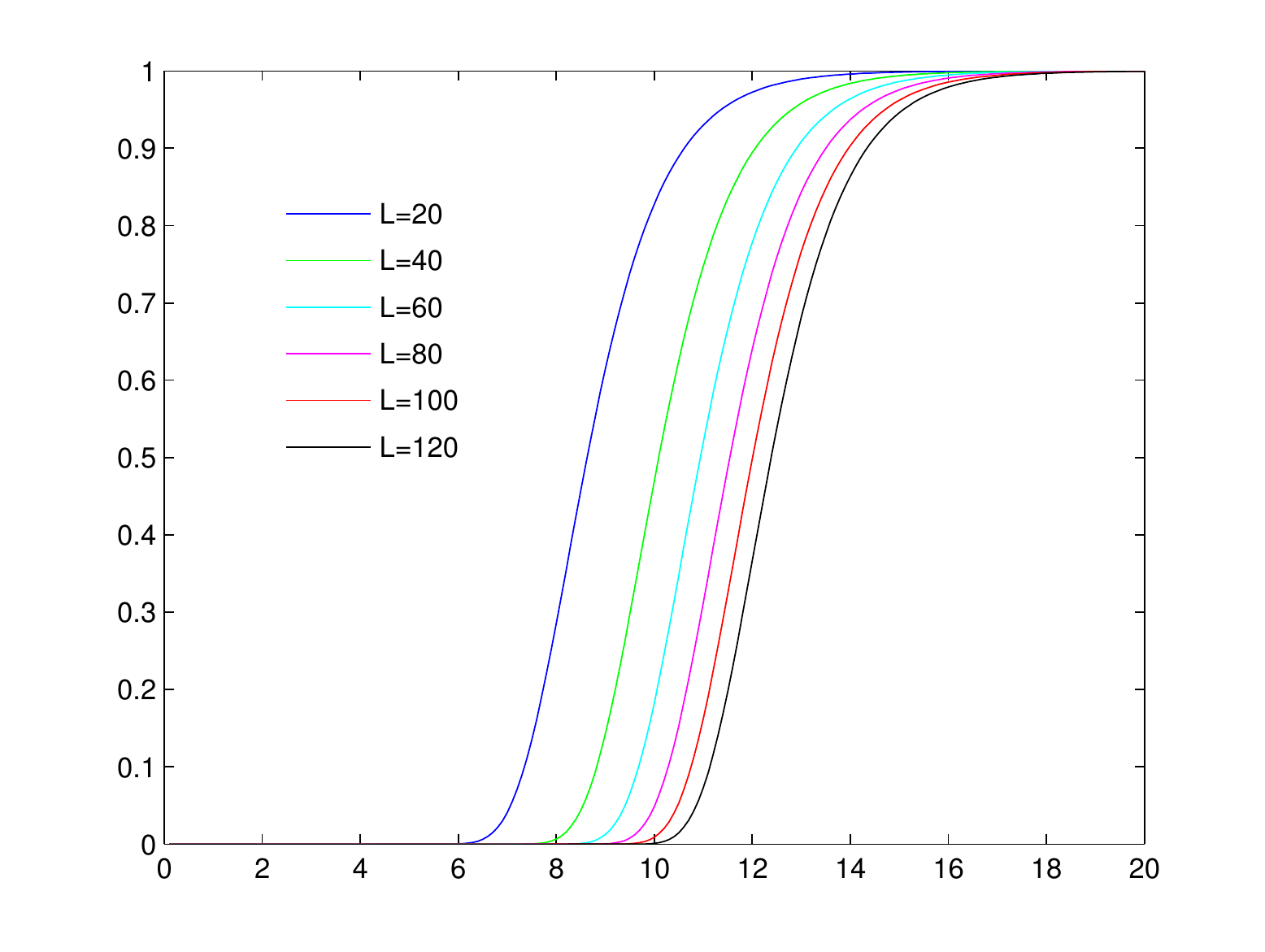}
\includegraphics[width=0.49\textwidth]{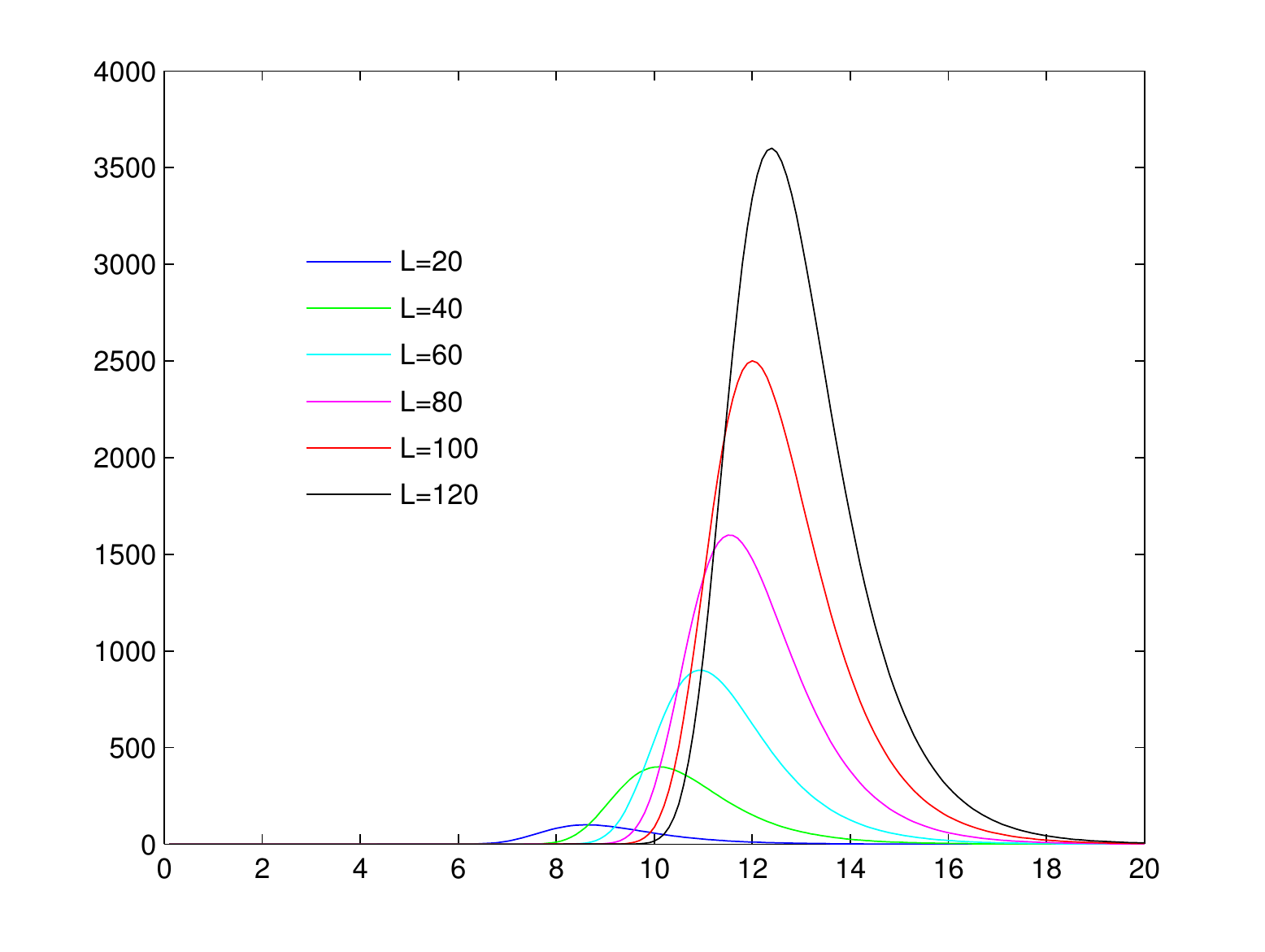}
\end{center}
\caption{Order parameter $z$ in $D=2$ (left) and its susceptibility $\chi$ 
(right).}
\label{fig_2}
\end{figure}

The $D =2$ case offers an interesting cross-check of the numerical results in 
higher dimensions, since
here everything can be calculated analytically. For the order parameter
$z$ and its susceptibility $\chi$ we have, for fixed volume $L^2$:
\begin{eqnarray}
\langle {z} \rangle_L &=& e^{-4 L^2 p(\beta)}\,;\qquad
\chi_L = L^2 \left[e^{-4 L^2 p(\beta)}-e^{-8 L^2 p(\beta)}\right]\\
p(\beta) &=& \frac{1}{2}\left[1-\frac{L_1(\beta)}{I_1(\beta)}\right] 
=
\sqrt{\frac{2 \beta}{\pi}}e^{-\beta}(1+{\cal{O}}(\frac{1}{\beta}))\,,
\end{eqnarray}
where $L$ and $I$ denote the modified Struve and Bessel functions, 
respectively. 
Plotting the above functions (see Fig.~(\ref{fig_2})) we can clearly 
distinguish a ``strong''
coupling regime, where the topology is ill defined, and a ``weak'' coupling
one, where $z$ takes the correct value it should have in the continuum theory.
The finite size scaling analysis can be performed exactly, giving for the 
susceptibility peaks and the corresponding pseudo-critical coupling:
\begin{equation}
\beta_c(L) = \ln{L^2} + \frac{1}{2}\ln{\ln{L^2}} + {\cal{O}}(1)\,;\qquad
\chi_L(\beta_c(L)) = \frac{L^2}{4}\,.
\label{eq_fss_2}
\end{equation}
From the above equation it is easy to extract the critical behaviour
of the correlation length around the critical coupling 
$\beta_c = \infty$, including logarithmic corrections:
\begin{equation}
\xi \sim \sqrt[4]{\frac{\pi \ln^2{2}}{2 \beta}}\;\cdot\, e^{\frac{1}{2}\beta}
\label{eq_ess}
\end{equation}
i.e. an essential scaling\footnote{Compare with the critical behaviour of the 
XY model, $\xi \sim \exp\left({b t^{-\nu}}\right)$ and $\chi \sim \xi^{2-\eta} 
\ln^{-2 r}{\xi}$, with $t = \left|T/T_c - 1\right|$ the
reduced temperature and $\nu = 1/2$, $\eta = 1/4$ 
$r = -1/16$ the critical exponents \cite{Kenna:1996bs}.} with critical 
exponents $\nu = 1$, $\eta = 0$ and $r=0$. 

Summarizing, although for any 
fixed volume $L^2$ one 
can always find a coupling above which the topology corresponds to that 
dictated by the boundary conditions, taking the thermodynamic 
limit first, as one should, the ``strong'' coupling regime extends to 
$\beta = \infty$ and the system is always in the disordered phase. 
A vortex topology cannot be defined.

Turning now to the $D =3$ case and using Monte-Carlo simulations to calculate 
$z$ we basically get the same picture. In Fig.~(\ref{fig_3}) we
show the susceptibility $\chi$ and its FSS with an essential scaling Ansatz,
again with $\beta_c = \infty$:
\begin{figure}
\begin{center}
\includegraphics[width=0.49\textwidth]{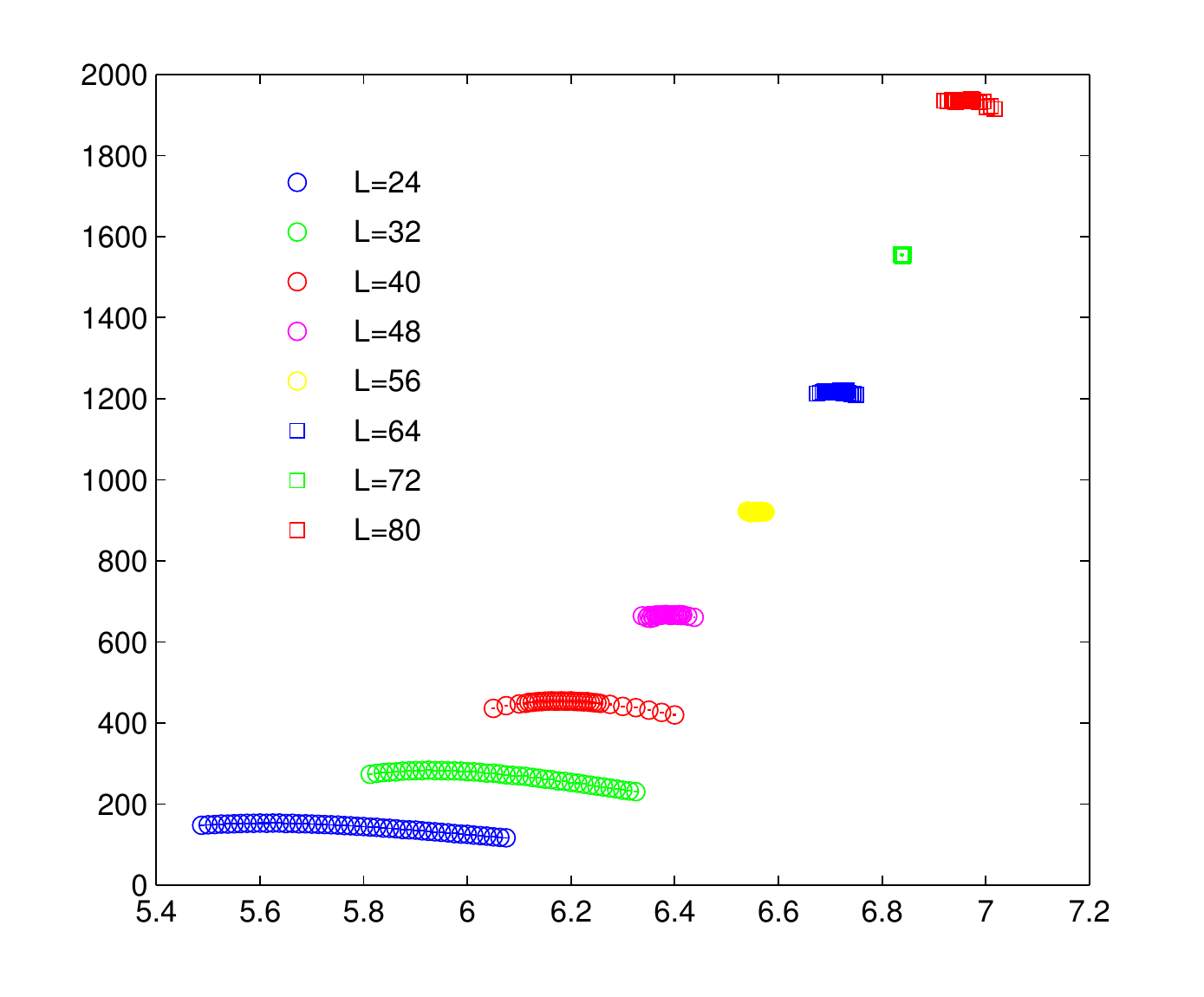}
\includegraphics[width=0.49\textwidth]{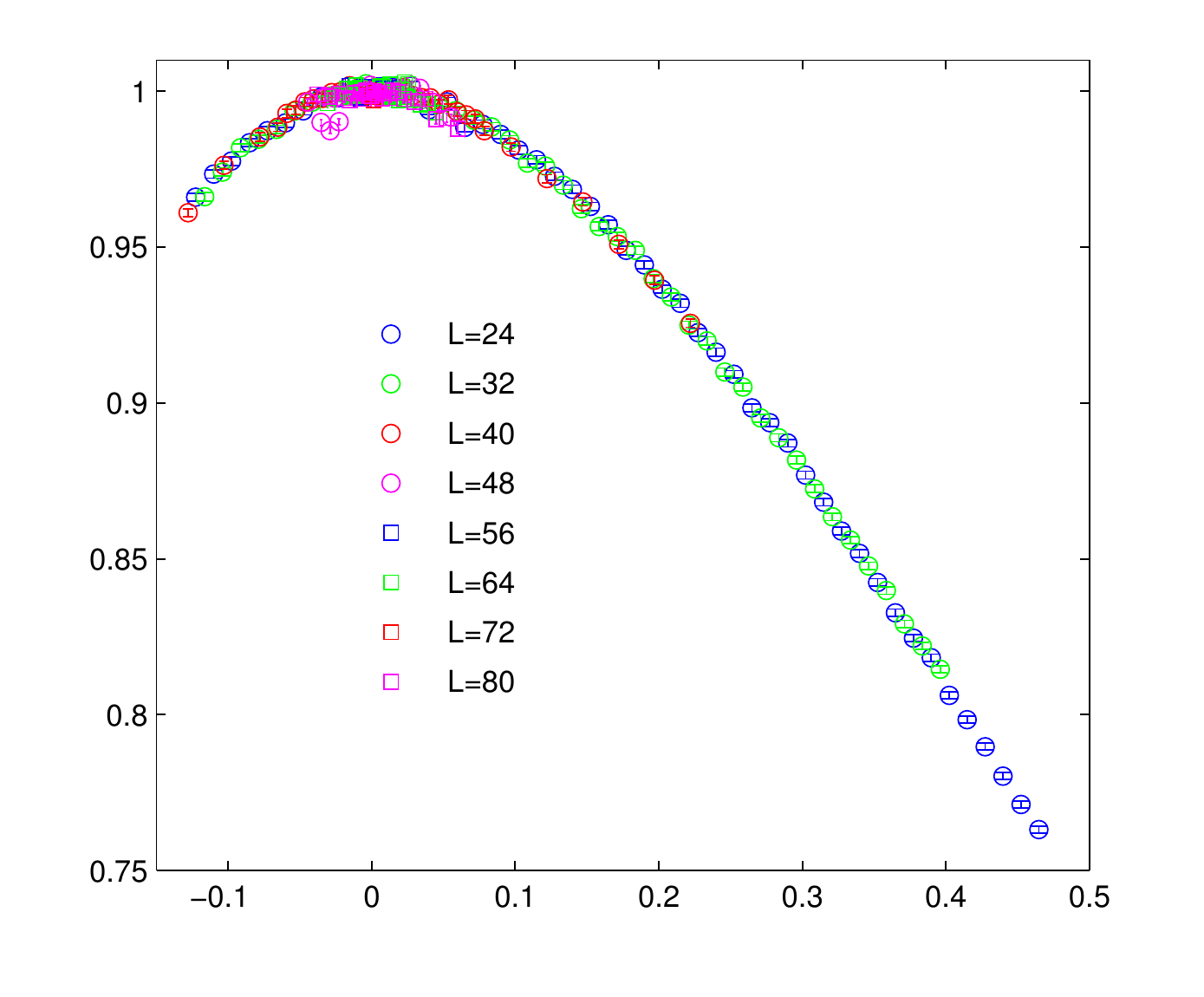}
\end{center}
\caption{Left: susceptibility $\chi$ of the order parameter $z$ in $D=3$. 
Right: same with FSS as in Eq.~(\protect\ref{eq_fss}).}
\label{fig_3}
\end{figure}
\begin{equation}
\beta_c(L)\simeq A \ln{L^2} + B\ln{\ln{L^2}}\,;\qquad\chi_L(\beta_c(L)) \simeq C {L^2} \ln^{-2 r}{L}
\label{eq_fss}
\end{equation}
The result is the same, i.e. in the continuum limit the theory is always 
in disordered phase and no vortex topology can be defined. The values
obtained for $\beta_c$ are well within the scaling region and for
fixed volumes $L^3$ they are always lower than the pseudo-critical coupling
at which the ``finite temperature'' transition for time length $L$
would be measured. 

\begin{figure}
\begin{center}
\includegraphics[width=0.49\textwidth]{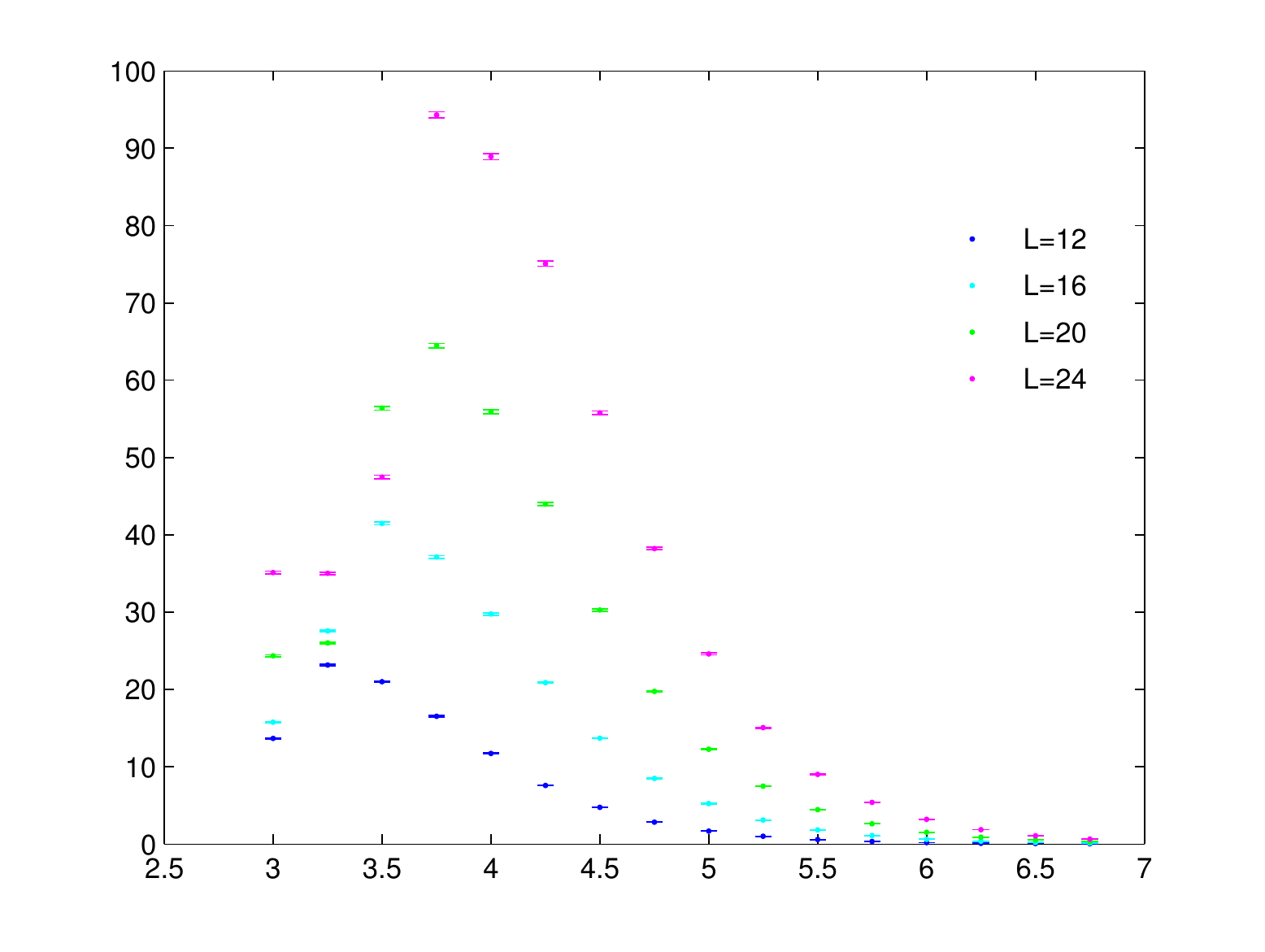}
\includegraphics[width=0.49\textwidth]{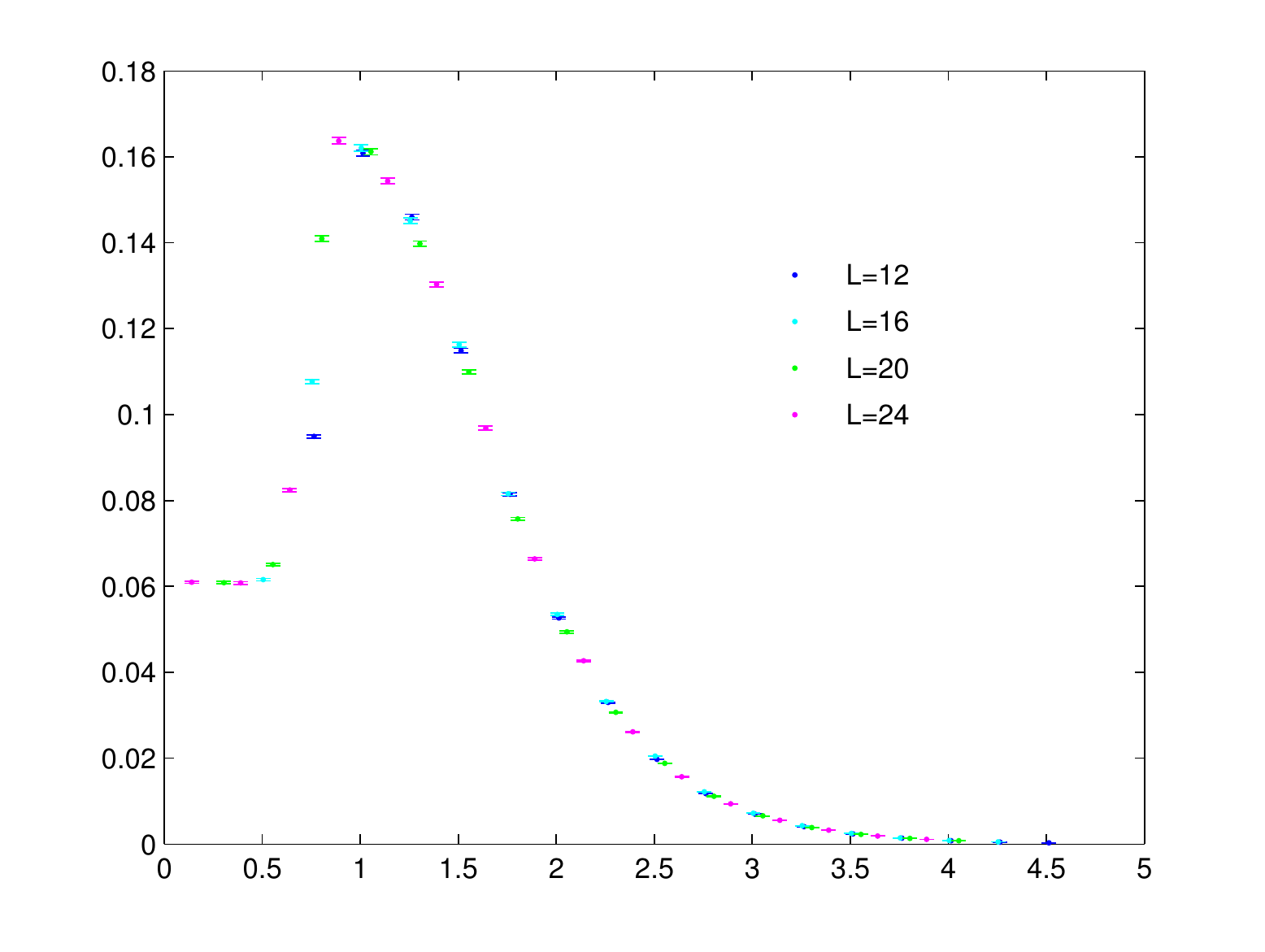}
\end{center}
\caption{Left: susceptibility $\chi$ of the order parameter $z$ in $D=4$. 
Right: same with FSS as in Eq.~(\protect\ref{eq_fss}).}
\label{fig_4}
\end{figure}

The situation is inverted at $D=4$. Here we still get the same result
as above for our order parameter $z$, i.e. an essential scaling as in 
Eq.~(\protect\ref{eq_fss}). The values of the pseudo-critical coupling 
$\beta_c$ are however higher than the values measured for the
deconfinement transition at time length $L$, cfr. Fig.~(\ref{fig_4}), making 
the scale at which topological vortices stabilize way above any physical
scale involved in the process. 
 
\section{Conclusions}

We have shown that $\forall\; D \leq 4$ the vortex topology for the 
standard Wilson action is always ill defined in the continuum limit.
This is driven by a too slow fall-off of discretization artefacts density, i.e. 
${\mathbb{Z}}_2$ magnetic monopoles Eq.~(\ref{eq:mon}). This means that for any 
fixed $\beta$ there always exists a lattice size $L$ for which enough open 
center vortices can form, spoiling the identification of topological sectors
and making a measurement of the conjectured super-selection rule in the 
thermodynamic limit ill-defined.

The details of such bulk effect will of course strongly depend on the 
discretization chosen. For example, the separation among the regimes in $D=3$ 
and $D=4$ are substantially different.
While in $D=3$ the deconfinement transition for fixed length $L$ always 
lies in the spurious phase above the pseudo-critical coupling $\beta_c(L)$ 
Eq.~(\protect\ref{eq_fss}), for $D=4$ the latter is always way above
the physical scale, making the physical volumes quite small.

Other choices for the discretization would of course change the picture.
For example, in $D=4$ one could define the theory through the
adjoint Wilson action, for which, above the the bulk 
transition \cite{Greensite:1981hw,Bhanot:1981eb,Halliday:1981te,%
Halliday:1981tm},
topology is well defined Ref.~\cite{Burgio:2006dc,Burgio:2006xj}. There
it was however found that $F\neq0$ in the confined phase, calling for
more investigations of the standard center vortices symmetry breaking argument 
as a model for confinement. Simulations in this case are however 
technically quite demanding. 

Another choice, which would in principle work in any dimensions, 
would be to define the discretized 
theory through a Positive Plaquette Model \cite{Fingberg:1994ut}. In this case 
the ${\mathbb{Z}}_2$ magnetic monopole constraint is always satisfied while 
the order parameter $z \equiv 1$ by construction. As
was proven in Ref.~\cite{Barresi:2006gq}, this is indeed
equivalent to simulating the adjoint Wilson action in a fixed topological 
sector. This model should then be the discretization of choice if one is 
interested in investigating the role of center vortices.

We would like to stress that none of the above results contradicts
universality. First, universality hold as long as no discretization artifacts
obstacle the continuum limit. In our case, center monopoles and the associated
open center vortices spoil the equivalence among different discretizations.
Second, all physical 
properties measurable in ``experiments'', like glueball masses and the 
critical exponents at the transition, should be reflected by physical 
observables which can be defined {\it irrespective} of the discretization 
chosen; this however does not mean that everything that can be defined in a 
given discretization should acquire a physical meaning.

\section*{Aknowledgements}
We wish to thank F. Bursa for useful discussions on the $D=2$ case. 
This work was partially supported the DFG under the contract DFG-Re856/6-3.

\end{document}